\documentclass[preprint,showpacs,preprintnumbers,amsmath,amssymb]{revtex4}
\tighten
\usepackage{graphicx}
\usepackage{dcolumn}
\usepackage{bm}
\begin{document}
\begin{titlepage}
\preprint{JLAB-THY-02-15}
\title{ The decay $\pi^0 \to \gamma\gamma$ to next to\\  leading order in Chiral Perturbation Theory}
\author{J.~L. Goity $^a$, A.~M. Bernstein $^{b}$, and B.~R. Holstein $^{c}$\\
$^a$ Department of Physics, Hampton University, Hampton, VA 23668, and\\
Thomas Jefferson National Accelerator Facility,\\ Newport News, VA
23606.\\
$^b$ Laboratory for Nuclear Science,  Massachusetts Institute of
Technology, \\
Cambridge, MA 02139.\\
$^c$ Department of Physics-LGRT,  University of Massachusetts, \\ Amherst, MA 01003.
 }

\begin{abstract}
The $\pi^0\to \gamma\gamma$ decay width is analyzed within the combined framework 
of Chiral Perturbation Theory and the $1/N_c$ 
expansion up to ${\cal{O}}(p^6)$ and
${\cal{O}}(p^4\times 1/N_c)$ in the decay amplitude. The $\eta'$ is explicitly included in the analysis. 
It is found that the decay width 
is enhanced by about 4.5\% due to the isospin-breaking induced mixing of the
 pure $U(3)$ states. This effect, which is of leading order in the low energy expansion,
 is shown to  persist nearly unchanged at next to leading order.
The chief prediction with its estimated uncertainty is 
$\Gamma_{\pi^0\to\gamma\gamma}=8.10\pm 0.08~ {\rm eV}$. 
This prediction at the 1\% level makes the upcomming precision measurement of the decay width even more urgent.  
  Observations on the $\eta$ and $\eta'$ can also be
made, especially about their mixing, which is shown to be significantly
affected by  next to leading order corrections.
\end{abstract}
\pacs{11.30.Rd, 11.40.-q, 12.39.Fe, 13.20.Cz}
\maketitle 
\medskip
\vfill
\end{titlepage}
\setcounter{page}{1}

\section{Introduction}

In the chiral SU(2) limit ($m_{u,d}=0$) the $\pi^0\to \gamma\gamma$
decay amplitude is precisely known to order $\alpha$\cite{dgh}. The
amplitude is ${\cal{O}}(p^4)$ in the low energy chiral counting.  It is 
determined entirely by the anomaly induced on the divergence of the
axial current $A_\mu^3=\bar{q}\gamma_\mu \gamma_5  \tau_3 q$ by the 
electromagnetic interaction,
and is expressed in terms of the only two available quantities---the
fine structure constant $\alpha$
and the pion decay constant $F_\pi$---the 
decay width in this limit is thus given 
by $\Gamma_{\pi^0\to \gamma\gamma}=(\frac{\alpha}{F_\pi})^2 
(\frac{M_{\pi^0}}{4 \pi})^3$.  The explicit breaking of 
chiral $SU_L(2)\times SU_R(2)$  symmetry
induced by non-zero u- and d-quark masses generates corrections to the
chiral limit result, and it is the purpose of the present work to evaluate 
these corrections as well as to understand their origin. In order to
achieve this goal it
it is crucial  to perform  the analysis in the extended framework
of  three flavors supplemented by the $1/N_c$ expansion 
in order to include explicitly the $\eta'$ degree of freedom.
  In this three flavor  framework the corrections turn out to be of two
types: 
\begin{itemize}
\item [i)] those due to 
isospin breaking (mixing corrections) that are proportional to $(m_u-m_d)/m_s$
or to $N_c (m_u-m_d)/\Lambda_{\chi}$, both giving contributions to 
the decay  amplitude that,  according to the counting  defined in the 
next section, are 
${\cal{O}}(p^4)$, {\it i.e.},  of the same order as the leading term,
and 
\item [ii)] those proportional to $m_{u,d}/\Lambda_{\chi}$ that
are of subleading order---${\cal{O}}(p^6)$---and which stem from 
different sources, as shown below. 
\end{itemize}
The inclusion of such corrections is crucial for a prediction of 
the $\pi^0 \to \gamma \gamma$ width  at the $ 1\%$ level, which is the
level of theoretical precision required  by  the forthcoming 
dramatic improvement expected in the experimental measurement of the
$\pi^0$ width via the Primakoff effect.  The PRIMEX 
experiment at Jefferson Lab \cite{PRIMEX}
is aiming at a measurement with an error about $1.5\%$, which is 
several times smaller 
than the $7.1\%$ uncertainty in the current world-average 
value $7.74\pm 0.55  \;{\rm eV}$
\cite{pdg}.  However,
this quoted experimental uncertainty is open to question as can be seen 
by the large dispersion of the experimental results which suggests
that the errors of the individual experiments have been underestimated.  Indeed, a direct measurement 
gives  $ \Gamma_{\pi^0}=7.25\pm 0.23\; {\rm eV}$\cite{direct},
a production experiment in $e^+-e^-$ collisions yields
$\Gamma_{\pi^0}=7.74\pm 0.66 \;{\rm eV}$\cite{Xball}, while Primakoff effect
experiments, all dating back to the early seventies, give disparate
values:  a large number---$11.61\pm 0.55 \; {\rm eV}$
\cite{Bellettini}, and two that are consistent with the current world 
average---$7.22\pm 0.55\; {\rm eV}$ \cite{Kryshkin}
and $7.93\pm 0.39\; {\rm eV}$ \cite{Browman}. This  unsatisfactory
experimental situation together with the rather precise theoretical 
prediction derived in the present work clearly lend great significance
to the upcoming PRIMEX measurement.  

Within the two-flavor framework, wherein the strange quark is integrated out,
any corrections to the amplitude are ${\cal{O}}(p^6)$ and reside
entirely in $F_{\pi^0}$ \cite{Bijnens} or in the ${\cal{O}}(p^6)$
odd-intrinsic parity Chiral Lagrangian\cite{p6WZ}, 
also  known as the ${\cal{O}}(p^6)$ Wess-Zumino (WZ) Lagrangian. At
leading order, the ensuing theoretical prediction (taking
$F_{\pi^0}=F_{\pi^+}$) is $\Gamma_{\pi^0\to \gamma\gamma}=7.725\;{\rm
eV}$, a result that agrees  well with the experimental world average within its
generous error.  The  analysis within $SU(2)$, however, does not
provide insight on the origin of the ${\cal O}(p^6)$ WZ  contribution just mentioned.
Such an insight {\it can} be gained by instead performing 
the analysis in the {\it three}-flavor framework, as has been 
shown by Moussallam\cite{Moussallam}. 
In particular, he pointed out that the primary corrections 
to the $\pi^0$ width result from the leading
order isospin breaking effects mentioned above in i), which stem 
from the $m_u\neq m_d$-induced mixing
between the pure isospin state  $\pi^0$ and pure $SU(3)$ states $\eta$ and $\eta'$ (to
be denoted below by $\pi_3$, $\pi_8$ and $\pi_0$ respectively).  
The present work confirms 
that such isospin breaking corrections persist as the dominant effect 
when next to leading order (NLO) corrections are included.

In this work then the 
$\pi^0\rightarrow\gamma\gamma$ decay rate is evaluated to NLO
 within $U_L(3)\times U_R(3)$ chiral perturbation theory wherein the $\eta'$ meson
is included consistently by means of the $1/N_c$ expansion since in
the large $N_c$ limit 
the  $\eta'$ becomes a Goldstone Boson.  Such a framework was recently developed
by Herrera-Sikl\'ody, Latorre, Pascual and Taron\cite{Herrera} and by 
Kaiser and Leutwyler\cite{Kaiser1,Kaiser2}, who showed that a 
simultaneous chiral and $1/N_c$ expansion leads to an
effective theory for the pseudoscalar nonet that is not only
internally consistent but is also very useful in practice, as the present work shows. 

The chief result of our paper is that the $\pi^0\rightarrow\gamma\gamma$
width is enhanced by about $4.5\%$ from the lowest order chiral anomaly
prediction, a result expected to hold within an uncertainty
of $~\pm 1\%$ after NLO contributions are included. The magnitude of this
enhancement agrees with that obtained in the analysis of
Moussallam \cite{Moussallam}, where the NLO corrections were not 
implemented in a consistent fashion, as done in the present work. 


\section{Two-photon decay amplitudes}

The decay amplitudes of $\pi^0$, $\eta$ and $\eta'$ into two photons 
can be obtained from the Ward identities satisfied by the three  axial vector currents $A_\mu^a={1\over 2}\bar{q}\gamma_\mu \gamma_5  \lambda^a q$ ($a=3,8,0$), where $\lambda^a$ are $U(3)$  generators ($\lambda^0$ being the $U(1)$ generator)  normalized via $\langle\lambda^a
\lambda^b\rangle\equiv
Tr(\lambda^a \lambda^b)=2 \delta^{ab}$.  In the presence of the strong
and electromagnetic
interactions, the divergence of the axial vector current is given by:
\begin{eqnarray}\label{div1}
\partial^\mu A_\mu^a&=&\frac{\alpha N_c}{4 \pi} \langle\lambda^a Q^2\rangle F\tilde{F} 
+\frac{\alpha_s}{4 \pi}\langle\lambda^a\rangle G\tilde{G}
+{i\over 2} \bar{q}\gamma_5  \{\lambda^a, {\cal{M}}_q\} q+\cdots,
\end{eqnarray}
where ${\cal{M}}_q$ is the quark mass matrix and $e Q$ is the electric
charge operator.  Here $F\tilde{F}=\frac{1}{2}
\epsilon_{\mu\nu\rho\sigma}
F^{\mu\nu}F^{\rho\sigma}$, $F_{\mu\nu}$ being the electromagnetic
field tensor, and similarly  
$G\tilde{G}=\frac{1}{2} \epsilon_{\mu\nu\rho\sigma}G^{a\mu\nu}
G^{a\rho\sigma}$, $G_{\mu\nu}$ being the gluon field \footnotemark[1]
\footnotetext[1]{Throughout the conventions in Bjorken and Drell are
used.}, and 
the ellipsis denotes terms irrelevant to this work. 

The two-photon amplitudes can  be obtained by considering the matrix
elements 
\begin{eqnarray}\label{div2}
\langle \gamma\gamma\mid \partial^\mu A_\mu^a \mid 0\rangle&=&
C_a \frac{\alpha N_c}{12 \pi}
\langle\gamma\gamma\mid  F\tilde{F}\mid 0\rangle+ \delta_{a 0} 
\frac{\sqrt{6}\alpha_s}{4 \pi} \langle\gamma\gamma\mid G\tilde{G} \mid 0\rangle\nonumber \\
&+& \frac{i}{2}  \langle\gamma\gamma\mid \bar{q} \gamma_5  \{ \lambda^a, {\cal{M}}_q\}
 q\mid 0\rangle,
\end{eqnarray} 
where  $C_3=1$,  $C_8=1/\sqrt{3}$, and $C_0=\sqrt{8/3}$.

If $p$ denotes the total momentum of the final two-photon state, then
in the limit of small $p^2$ Eq. \ref{div2} admits a low energy
expansion and can be expressed as:
\begin{eqnarray}\label{div3}
&& \sum_{\bar{a}} 
\langle\gamma\gamma\mid \pi_{\bar{a}},p\rangle
\langle\pi_{\bar{a}},p\mid \partial^\mu A_\mu^a\mid 0\rangle
\frac{i}{p^2-M_{\bar{a}}^2+i 
\epsilon}=C_a \frac{\alpha N_c}{12 \pi}
\langle\gamma\gamma\mid  F\tilde{F}\mid 0\rangle\nonumber\\
&+& \delta_{a 0} 
\frac{\sqrt{6}\alpha_s}{4 \pi}
\sum_{\bar{a}} \langle\gamma\gamma\mid \pi_{\bar{a}},p\rangle
\langle\pi_{\bar{a}},p\mid G\tilde{G} \mid 0\rangle \frac{i}{p^2-M_{\bar{a}}^2+i \epsilon}
\nonumber \\
&+& \frac{i}{2}
\sum_{\bar{a}} \langle\gamma\gamma \mid \pi_{\bar{a}},p\rangle
\langle\pi_{\bar{a}},p\mid \bar{q} \gamma_5  \{ \lambda^a, {\cal{M}}_q\}
q\mid 0\rangle 
\frac{i}{p^2-M_{\bar{a}}^2+i \epsilon}+\cdots ,
\end{eqnarray}
where $\langle\gamma\gamma\mid \pi_{\bar{a}},p\rangle $ are the two-photon 
amplitudes, the ellipsis denotes contributions from excited mesons as well
as from the continuum, all being of NLO or higher, and 
the mass eigenstates $\pi_{\bar{a}}$ that correspond to the physical 
$\pi^0$, $\eta$ and $\eta'$
are given by
\begin{equation}\label{mixing}
 \pi_{\bar{a}}=\sum_{a=3,8,0} \Theta_{\bar{a} a} \pi_a~~,
\end{equation}
 where the mixing matrix that diagonalizes the mass matrix is parametrized in terms of Euler angles
$\theta_3$, $\theta_8$ and $-\theta_0$
$$\Theta= \left(
 \begin{array}{ccc} c_3 c_8-c_0 s_3 s_8 &  c_3 s_8+ c_8 c_0 s_3 & -s_3 s_0\\
  -c_8 s_3-c_3 c_0 s_8 & -s_3 s_8+c_3 c_8 c_0 & -c_3 s_0 \\
    - s_8 s_0  & c_8 s_0   & c_0 
 \end{array} \right),$$ 
where $c_i=\cos \theta_i$ and $s_i=\sin \theta_i$.  Here, for small mixing, the
projection of the physical $\pi^0$ onto $\pi_8$ is given by the angle 
$ \epsilon\simeq \theta_3+\theta_8$,
the projection of the physical $\eta$ onto $\pi_0$ is 
approximately given by  $-\theta_0$ ($\theta_0$ can 
therefore be identified with the well known  $\eta-\eta'$ mixing angle as it is customarily defined), and
the projection of the  physical $\pi^0$ onto $\pi_0$  is given by  $\tilde{\epsilon}\simeq -\theta_3  \theta_0$.

The NLO---{\it i.e.}, ${\cal{O}}(p^6)$---corrections in Eq. \ref{div3} 
reside in the terms displayed explicitly through their dependence 
on the masses and decay constants, as well as in 
pieces that stem from continuum and excited states. In works that preceded 
that of Moussallam \cite{Moussallam}, such as
references\cite{Bijnens} and \cite{dh}, 
such mixing corrections 
as well as the NLO effects of the latter kind  were disregarded. 
Ignoring such effects
implies that only  NLO corrections which are absorbed into
the $\pi^0$ decay constant remain, as it was shown in  \cite{Bijnens}; 
in that case and  taking $F_{\pi^0}=F_{\pi^+}$, the predicted $\pi^0$
width is the previously mentioned $7.725\;{\rm eV}$. As shown here,
however, disregarding mixing in particular constitutes a very  poor approximation. 
In the presence of mixing  the pseudoscalar decay constants form a 
$3\times 3$ matrix defined 
by the matrix elements of the axial vector currents, which connect the
pseudoscalar mesons to the vacuum:
\begin{equation}
\langle\pi_{\bar{a} },p\mid A_\mu^a \mid 0 \rangle=-i p_\mu F_{\bar{a} a}.
\end{equation}
Indeed  this decay-constant matrix contains all that is needed to calculate
the three two-photon amplitudes, except the contributions stemming from the ${\cal{O}}(p^4)$ WZ Lagrangian,  
and its evaluation is the centerpiece of the present work.

As already mentioned in the introduction, the  present analysis includes the  $\eta'$ as an explicit degree of freedom, 
which, in order to be  consistently implemented in an effective theory, requires the
validity of the $1/N_c$ expansion, wherein, taking the chiral 
$SU_L(3)\times SU_R(3)$ limit, $M_{\eta'}^2={\cal{O}}(1/N_c)$. Thus, in the
framework of the $1/N_c$ expansion $M_{\eta'}^2$ should be considered as ``small,''
and its explicit inclusion becomes consistent with having a simultaneous
low energy chiral expansion. This explicit inclusion of the $\eta'$ in
the low energy expansion implies that $M_{\eta'}^2$ must count as a 
quantity of ${\cal{O}}(p^2)$, which in turn  implies that $1/N_c$
should be counted as a quantity of the same order.
 Indeed, a consistent effective theory can be formulated with such a
counting scheme\cite{Herrera,Kaiser1,Kaiser2}, and it is interesting
to note that  $1/N_c$ and the magnitude of $SU(3)$ breaking are
comparable in size in the real world.

Taking the chiral limit---${\cal M}_q\rightarrow 0$---in 
Eq. \ref{div3} and neglecting the electromagnetic
piece---{\it i.e.}, the $F\tilde{F}$ term---, equating the residues at
$p^2=M_{\eta'}^2$ lead to the well known relation 
\begin{equation}\label{Metap}
M_0^2=\sqrt{6}\frac{\alpha_s}{4 \pi F_0} \langle\eta'|G\tilde{G}\mid 0\rangle,
\end{equation}
with $M_0$ being the $\eta'$ mass in the chiral limit.  Here the lowest
order result 
$$\langle\pi_{\bar{a}}, p\mid\partial^\mu A_\mu^a\mid 0\rangle
=\delta_{\bar{a}a}p^2 F_0$$ 
was used, where $F_0$ is the pion decay
constant in the chiral limit. In the large $N_c$ limit, $F_0$ scales as
$\sqrt{N_c}$, while in the chiral limit the ratio $F_0/F_{\eta'}$ is equal
to unity up to corrections of order $1/N_c$.

On the other hand, for non-vanishing quark masses equating the
residues in Eq. \ref{div3} yields
\begin{eqnarray}\label{pseudoscalar}
\langle\pi_{\bar{a}}, p\mid  \bar{q} \gamma_5  \{ \lambda^a, {\cal{M}}_q\} q \mid 0\rangle&=&
-2 i (\langle\pi_{\bar{a}}, p\mid \partial^\mu A_\mu^a \mid 0\rangle\nonumber\\
&-&\delta_{a0}
\sqrt{6}\frac{\alpha_s}{4\pi} \langle\pi_{\bar{a}},p \mid G\tilde{G}\mid 0\rangle)
\end {eqnarray}
where $p^2=M_{\bar{a}}^2$. As is well known, the $p^2$ dependence of the LHS 
appears first at ${\cal{O}}(p^6)$ \cite{GandL1}, which would affect the two-photon amplitudes at ${\cal{O}}(p^8)$, {\it i.e.},   beyond the
accuracy needed in this  work. 
Thus, it is consistent to use Eq. \ref{pseudoscalar} at $p^2=M_{\bar{a}}^2$ to represent the LHS in the entire low $p^2$ domain. 

At LO---${\cal{O}}(p^4)$--- then, 
Eqns. \ref{div3}, \ref{Metap}, and \ref{pseudoscalar} yield immediately the result for the two-photon amplitudes
\begin{equation} \label{LOamplitudes}
\langle\gamma\gamma\mid \pi_{\bar{a}}\rangle =\sum_{\bar{a=3,8,0}}
-i\frac{\alpha N_c}{12 \pi} C_a F^{-1}_{a \bar{a}} \langle\gamma \gamma\mid F \tilde{F}\mid 0\rangle
\end{equation}
At this order the decay constant matrix is simply  given by
\begin{equation}\label{FatLO}
F_{\bar{a}a}=\Theta_{\bar{a}a} F_0
\end{equation}
where  $\Theta_{\bar{a}a}$ is the mixing matrix obtained from the 
${\cal{O}}(p^2)$ mass formulas.
Of course, the result of Eq. \ref{LOamplitudes} coincides with the result obtained by
means of the ${\cal{O}}(p^4)$ WZ term including explicitly the singlet pseudoscalar. The purpose
of carrying out the above Ward identity analysis is, however, to make more
transparent the origin and structure of the  higher order corrections
and to set the stage for the inclusion of NLO modifications.

\section{Leading Order Results}

The leading order mass formulas are obtained from the ${\cal{O}}(p^2)$
chiral Lagrangian. The $U(3)$ field is parametrized by the unitary matrix:
\begin{equation}\label{Ufield}
U=\exp\left({i \sum_{a=0}^{8}\frac{\pi_a \lambda^a}{F_0}}\right)
\end{equation}
where $F_0=92.42~{\rm MeV}$ at LO. The ${\cal{O}}(p^2)$  Lagrangian with the standard definitions  of  covariant derivatives and  sources $\chi
$ \cite{GandL1} is given by:
\begin{eqnarray}\label{L2}
{\cal{L}}^{(2)}&=&\frac{1}{4} F_0^2 \langle D_\mu U D^\mu U^{\dagger}\rangle 
+ \frac{1}{4} F_0^2 \langle\chi U^{\dagger}+\chi^{\dagger} U\rangle-\frac{1}{2} M_0^2 \pi_0^2.
\end{eqnarray}
and the mass matrix in the $\pi_3$, $\pi_8$, $\pi_0$ sector of interest
resulting from ${\cal{L}}^{(2)}$ is:
\begin{equation}\label{LOmass}
M^2_{LO}=B_0\left(
 \begin{array}{ccc} 
2  \hat{m} & \frac{1}{\sqrt{3}} (m_u-m_d)  &  \sqrt{\frac{2}{3}}  (m_u-m_d) \\
\frac{1}{\sqrt{3}} (m_u-m_d)   & \frac{2}{3}  (\hat{m}+2 m_s)  & -\frac{\sqrt{8}
}{3}  (m_s-\hat{m}) \\
 \sqrt{\frac{2}{3}}  (m_u-m_d)  &   -\frac{\sqrt{8}}{3}  (m_s-\hat{m}) &~~~\frac{M_0^2}
{B_0}+\frac{2}{3}  (2 \hat{m} +m_s).
 \end{array} \right)
\end{equation}
Using the leading order mass formulas---{\it e.g.},
$M_{\pi^+}^2=2B_0 \hat{m}$, with $2\hat{m}=m_u+m_d$---and 
extracting isospin breaking from the $K^+-K^0$ mass difference via Dashen's 
theorem to eliminate the EM contributions\footnotemark[3]
\footnotetext[3]{Throughout, the meson masses used are those with EM
contributions subtracted.}, a best fit to the masses yields a 
singlet mass $M_0$ of approximately 850 MeV, and the Euler angles: $\theta_3=1.57^o$, 
$\theta_8=-0.56 ^o$, and $\theta_0=-18.6 ^o$. This fit yields then
 $\epsilon=\theta_3+\theta_0\sim 1^o$, which is substantially larger
than the value  \cite{GandL1}
\begin{equation}\label{SU3mixing}
\epsilon={\sqrt{3}\over 4}{m_d-m_u\over m_s}\simeq 0.56^o
\end{equation}
that arises in the  limit $M_0\to \infty$, when only octet degrees of 
freedom are included. The LO mass matrix, however, gives a poor result
for the masses. In particular the $\eta$ mass is too low by almost
50 MeV, a problem that is generic at LO in the low energy and $1/N_c$ 
expansions\cite{Georgi}.

Using the relations 
\begin{equation}\label{genampl}
\langle\gamma\gamma \mid \pi_{\bar{a}}\rangle=\kappa_{\bar{a}} 
\langle\gamma\gamma \mid F\tilde{F} \mid 0\rangle,~~~~~\Gamma_{\bar{a}}=\mid \kappa_{\bar{a}
}\mid ^2\frac{M_{\bar{a}}^3}{4 \pi}
\end{equation}
which connect the decay amplitudes and associated widths, 
 the fitted parameters at LO lead to a $\pi^0\to \gamma\gamma$ decay width 
of 8.08 eV, which is $4.5\%$  larger than the  leading order
result $\Gamma_{\pi^0\to \gamma\gamma}=7.725~{\rm eV}$ obtained in
the two-flavor framework wherein mixing effects are moved to NLO.  It
should be noted, however, that the two-photon widths of the $\eta$ and
$\eta'$ predicted in this leading order fit are too large, the first
being  22\% and the second 20\% larger than the corresponding
experimental values.  One of the chief reasons  for this disagreement is
 that  $SU(3)$ breaking in the pseudoscalar decay constants is not included at LO.
 The $\eta-\eta'$  mixing angle  $\theta_0$ turns out to be $\sim -18.6^o$ 
at LO, and will be reduced by almost a factor of two when NLO corrections are included.
The magnitude of the observed  LO enhancement of the $\pi^0$ width 
is in line with the ratio of 
isospin breaking versus $m_s$ ($(m_d-m_u)/m_s\simeq 2.3\%$)
and versus $M_0$ ($B_0 (m_d-m_u)/M_0^2\simeq 1.1\%$), and is therefore
not surprising.  It is important to note that the corrections due to
mixing with $\eta$ and with $\eta'$ are of the same sign and of similar
magnitude. This point is the primary reason why the  $\eta'$ {\it
must} be  explicitly included for a full understanding of the mixing
effects. Although $\eta'-\pi^0$ mixing is smaller than the $\eta-\pi^0$ mixing,
the bare singlet state has an intrinsic two-photon amplitude larger by a factor 
$\sqrt{8}$ that  compensates for the smaller mixing.  Overall,
however, the LO fit is poor, and
dramatic improvement results when  the NLO corrections are included,
as shown in the following section.

\section{Next to leading order analysis and results}

At NLO the amplitudes receive corrections of two types---those that 
affect the decay constants
and mixing angles, and those that stem from the presence of excited states
 and which are included in the ${\cal{O}}(p^6)$ WZ  Lagrangian. 

The first type of correction requires the determination of the masses
 and decay constants to NLO and can be obtained in the standard fashion  
by calculating 
the  two-point functions of axial vector  currents, where the
relevant diagrams are shown in Fig. 1.  Up to ${\cal{O}}(p^2)$
and  ${\cal{O}}(p^0/N_c)$ such two-point functions require only the
effective Lagrangians ${\cal{L}}^{(2)}$  and  ${\cal{L}}^{(4)}$. 
Chiral loop corrections are ${\cal{O}}(p^2/N_c)={\cal{O}}(p^4)$, and therefore beyond
the precision of the present calculation, but such loop corrections 
will be calculated merely as a means to estimate the size of
possible contributions from terms of that order and also as a test on 
the practical validity 
of the $1/N_c$ expansion. The lowest order Lagrangian has
already been given in Eq. \ref{L2}, while the next to leading order 
Lagrangian ${\cal{L}}^{(4)}$ has the form given by \cite{GandL1,Herrera,Kaiser1,Kaiser2}
\begin{eqnarray}\label{L4}
{\cal{L}}^{(4)}&=&\cdots + 
L_4 \langle D_\mu U^{\dagger} D^\mu U\rangle\langle\chi^{\dagger} U+U^{\dagger} \chi\rangle  +
L_5 \langle D_\mu U^{\dagger} D^\mu U(\chi^{\dagger} U+U^{\dagger} \chi)\rangle\nonumber\\
&+&L_6 \langle\chi^{\dagger} U+U^{\dagger} \chi\rangle^2+ 
L_7 \langle\chi^{\dagger} U-U^{\dagger} \chi\rangle^2+   
L_8 \langle\chi U^\dagger \chi U^\dagger+h.
c.\rangle\nonumber\\ 
& +& \frac{\Lambda_1}{2} D_\mu \pi_0 D_\mu \pi_0
-\frac{i F_0 \Lambda_2}{2\sqrt{6}}  \pi_0 \langle\chi U^\dagger-\chi^\dagger U\rangle\nonumber\\
&+& i L_{18}\sqrt{6} D^\mu \pi_0 \langle   D^\mu U^{\dagger} \chi-D^\mu U \chi^{\dagger}\rangle+
i L_{25} \sqrt{6} \pi_0 \langle  U^{\dagger} \chi  U^{\dagger}\chi- U\chi^{\dagger} U \chi^{\dagger}\rangle+ \cdots,
\end{eqnarray}
where only the terms relevant to this work are included. In the presence of the $SU(3)$ singlet axial vector
source field $a_\mu^0$, $D_\mu \pi_0=\partial_\mu \pi_0-F_0 \, a_\mu^0$. At this point it is important to note that 
the singlet axial current has non-vanishing anomalous dimension \cite{Kodaira}, which implies that some 
low energy constants (LECs) as well as the singlet field $\pi_0$ must be renormalized  (the corresponding renormalized
quantities will thus depend on the QCD renormalization scale $\mu_{QCD}$) \cite{Kaiser1,Kaiser2}. 
Since the renormalization of the axial current is subleading in
$1/N_c$, such dependence appears first  at the level of 
${\cal{L}}^{(4)}$ LECs. It has been found in particular that the 
LECs $\Lambda_1$, $\Lambda_2$, $L_{18}$ and $L_{25}$, all of which are
 subleading in $1/N_c$, must be
renormalized\cite{Kaiser1}, implying that the values of these LECs will depend on the
value of the scale  $\mu_{QCD}$. Other quantities  such as
$F_{\eta'}$ and the 
singlet $\pi_0$ field must be renormalized as well and depend on 
$\mu_{QCD}$ through the renormalization factor $Z_A$ associated with the singlet axial current. It is very convenient to make use of the asymptotic freedom of QCD to set   $\mu_{QCD}$
arbitrarily large and give the values of the LECs in that limit. Indeed, the renormalization factor of the axial current $Z_A$
evolves to a fixed point that can be taken to be $Z_A=1$ as
$\mu_{QCD}\to \infty$. 
All quantities given in the following that have a dependence on
$\mu_{QCD}$ are then taken  in this limit.  It is well known that the 
low energy constants $L_5$ and $L_8$ are ${\cal O}(N_c)$, while $L_4$
and $L_6$ are subleading and ${\cal O}(N_c^0)$ \cite{GandL1}
 and are  needed in order to renormalize the one-loop 
contributions from ${\cal{L}}^{(2)}$. With $\eta'$ as an explicit degree of freedom, 
$L_7$ is also subleading in $1/N_c$.   The renormalized pieces of 
the subleading LECs  are therefore set to zero at the chosen chiral renormalization scale $\mu$  in our analysis. 
On the other hand, the LECs 
$\Lambda_1$ and $\Lambda_2$ are ${\cal O}(1/N_c)$ and the 
corresponding terms in the Lagrangian are ${\cal O}(p^2/N_c)={\cal O}(p^4)$ and 
thus must be included in the calculation.  The low energy constant
$\Lambda_1$ provides an ${\cal{O}}(1/N_c)$ correction to the 
$\eta'$ decay constant, while both $\Lambda_1$ and $\Lambda_2$ affect
entries in the mass matrix involving the $\eta'$ at order 
${\cal{O}}(p^2/N_c)$. Finally, the terms involving the  LECs $L_{18}$ and $L_{25}$ are of ${\cal{O}}(p^4/N_c)$ which is 
beyond the order of the present calculation and therefore their renormalized pieces are set to vanish as well.
The renormalized LECs are defined in the usual $\overline{MS}$
renormalization scheme\cite{GandL1}:
\begin{eqnarray}\label{LECsren}
L_i=L_i^r(\mu)+\Gamma_i \lambda(\mu) &~& \Lambda_i=\Lambda_i^r(\mu)+\Delta_i \lambda(\mu) \nonumber\\
\lambda(\mu)=\frac{\mu^{d-4}}{16 \pi^2}
\left(\frac{1}{d-4}\right. &\! -\! &\left. \frac{1}{2}(\log 4\pi+1+\Gamma'(1))\right)
\end{eqnarray}

The $\beta$-functions $\Gamma_i$ associated with the LECs $L_i$, and $\Delta_i$ associated with 
$\Lambda_i$ that result from the chiral one-loop calculation are given by \cite{Herrera,Kaiser1,Kaiser2}: 
$\Gamma_4=1/8$, $\Gamma_5=3/8$, $\Gamma_6=1/16$, $\Gamma_7=0$, $\Gamma_8=3/16$, 
$\Gamma_{18}=-1/4$,  $\Gamma_{25}=0$, $\Delta_1=-1/8$, and $\Delta_2=3/8$.

\begin{figure*}
\includegraphics{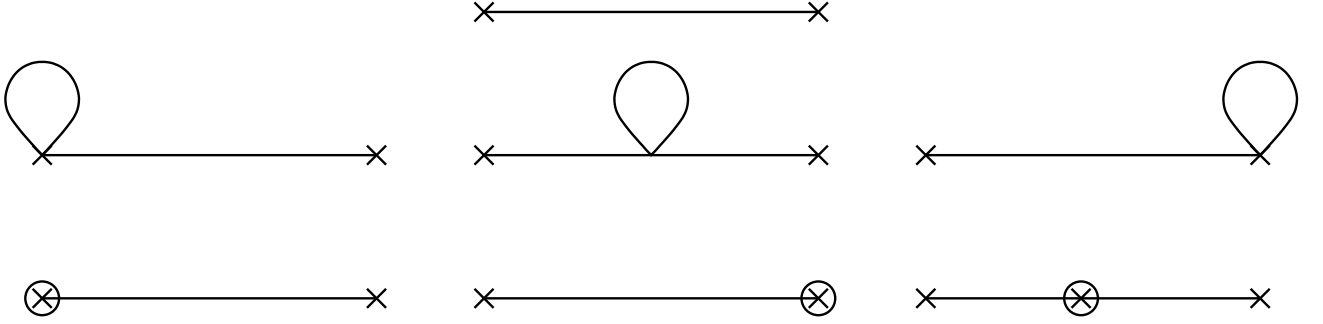}
\caption{\label{fig:wide} Two-point function of axial currents to NLO. The last three diagrams involve the countertem insertions from ${\cal{L}}^{(4)}$}
\end{figure*}


The two-point functions of axial currents can be written in momentum 
space as
\begin{equation}\label{twopoint}
\int d^4x \; e^{ip\cdot x}\langle 0\mid T( A_\mu^a(x) A_\nu^b(0)\mid 0 \rangle=p_\mu p_\nu 
\sum_{\bar{a}} F^T_{b \bar{a}}(p^2) 
\Delta_{\bar{a}}(p^2)
F_{\bar{a} a}(p^2)+\cdots,
\end{equation}
where the term explicitly shown contains the light pseudoscalar poles
and $\Delta_{\bar{a}}(p^2)$ is the propagator of the corresponding mass eigenstate. 
From  the location of the low energy poles and the residues of
the two-point function the light pseudoscalar decay constants and masses are
extracted, yielding 
\begin{eqnarray}\label{NLOmass}
M_{ab}^2=M_{LO\;ab}^2-(\sigma_{CT}+\sigma_{loop}-\frac{1}{2 F_0}
\left\{ M_{LO}^2, \phi_{CT}+\frac{1}{2}\phi\right\})_{ab},~~a,b=3,8,0\nonumber\\
M_a^2=M_{LO\;a}^2-((\sigma_{CT}+\sigma_{loop})_a-\frac{1}{ F_0}
M_{LO\; a}^2(\phi_{CT}+\frac{1}{2}\phi)_a), ~~a\neq 3,8,0
\end{eqnarray}
The corresponding decay constants are given by:
\begin{eqnarray}\label{NLOF}
F_{\bar{a}a}&=& \Theta_{\bar{a}b} F_{ba} \nonumber \\
 F_{ab}&=&F_{ba}= F_0\; 
(\delta_{ab}-\frac{1}{2}(\phi_{CT}+\frac{3}{2}\phi)_{ab}), ~~a,b=3,8,0\nonumber\\
F_{a}&=&F_0-\frac{1}{2}(\phi_{CT}+\frac{3}{2}\phi)_{a}, ~~a\neq 3,8,0
\end{eqnarray}
where the following definitions were made:
\begin{eqnarray}
\phi_{CT\;ab}&=&-\left(\frac{4 B_0 L_5^r(\mu)}{F_0}
\langle\{\lambda^a,\lambda^b\}{\cal{M}}_q\rangle+\Lambda_1 
\delta_{a0} \delta_{b0}\right)\nonumber\\
\phi_{ab}&=&-\frac{1}{12 F_0}\left(\sum_{c,d=3,8,0}\gamma^{abcd}
\Theta^T_{c\bar{a}}\mu_{\bar{a}}\Theta_{\bar{a}d}+
\sum_{c\neq 3,8,0}\gamma^{abcc}\mu_c\right)\nonumber\\
\sigma_{CT\;ab}&=&-\frac{8 B_0^2 L_8^r(\mu)}{F_0^2}
\langle \{{\cal{M}}_q,\lambda^a\}\{ {\cal{M}}_q, \lambda^b\}\rangle\nonumber\\
&-&2\sqrt{\frac{2}{3}} \Lambda_2 B_0
(\delta_{a0}\langle\lambda^b{\cal{M}}_q\rangle+
\delta_{b0}\langle\lambda^a{\cal{M}}_q\rangle)\nonumber\\
\sigma_{ab}&=&\frac{1}{24 F_0^2}\left(\sum_{c,d=3,8}
\gamma^{abcd}\Theta^T_{d\bar{a}}\mu_{\bar{a}} M_{\bar{a}}^2 \Theta_{\bar{a}d}+
\sum_{c\neq 3,8}\gamma^{abcc}\mu_c M_c^2\right)\nonumber\\
&+&\frac{B_0}{24 F_0^2}\left(\sum_{c,d=3,8,0}
{\cal{M}}^{abcd}\Theta^T_{c\bar{a}}\mu_{\bar{a}}\Theta_{\bar{a}d}+
\sum_{c\neq 3,8,0} {\cal{M}}^{abcc} \mu_c\right),
\end{eqnarray}
and,
\begin{eqnarray}
\mu_{\bar{a}}&=&\frac{1}{16 \pi^2}M_{\bar{a}}^2 \log\frac{M_{\bar{a}}^2}{\mu^2}\nonumber\\
\gamma^{abcd}&=&\langle[\lambda^a,\lambda^c][\lambda^b,\lambda^d]\rangle\nonumber\\
{\cal{M}}^{abcd}&=&\frac{1}{2}\sum_{perm.\{\sigma\}}
\langle{\cal{M}}_q\lambda^{\sigma_a}
\lambda^{\sigma_b}\lambda^{\sigma_c}\lambda^{\sigma_d}\rangle.
\end{eqnarray}
Throughout, the terms whose renormalized LECs are set to vanish have not been displayed explicitly.
It is interesting to note that $\phi_{ab}$ does not receive any loop contributions from the singlet pseudoscalar mode, implying that $F_{ab}$ is also free of such contributions.


It is useful at this point to give the explicit expressions of the  masses 
and decay constants at NLO  disregarding the chiral logarithms. For
the decay constants the above expressions lead to
\begin{eqnarray}
F_{\pi^+}&=&F_0+\frac{4 L_5 B_0 }{F_0} (m_u+m_d)\nonumber\\
F_{K^+}&=&F_0+\frac{ 4 L_5 B_0 }{F_0} (m_u+m_s)\nonumber\\
F_{33}&=&F_{\pi^+}\nonumber\\
F_{88}&=&F_0+\frac{4 L_5 B_0}{3F_0}(m_u+m_d+4 m_s)\nonumber\\
F_{00}&=&F_0(1+\frac{\Lambda_1}{2})+\frac{8  L_5 B_0 }{3 F_0}
(m_u+m_d+ m_s)\nonumber\\
F_{38}&=& \frac{4  L_5 B_0 }{\sqrt{3}F_0}(m_u-m_d)
\nonumber\\
F_{30}&=&\sqrt{2} F_{38}\nonumber\\
F_{80}&=&\frac{\sqrt{32} L_5 B_0}{3 F_0} (m_u+m_d-2 m_s)
\end{eqnarray}
while for the masses, 
\begin{eqnarray}
M_{\pi^+}^2&=& 2 B_0 \hat{m}+\frac{32 (2L_8-L_5) }{F_0^2}    B_0^2 \hat{m}^2 
\nonumber\\
M_{K^+}^2&=&B_0(m_u+m_s) +\frac{8 (2L_8-L_5) }{F_0^2} B_0^2 (m_u+m_s)^2
\nonumber\\
M_{K^0}^2&=&B_0(m_d+m_s) +\frac{8 (2L_8-L_5) }{F_0^2} B_0^2 (m_d+m_s)^2
\nonumber\\
M_{33}^2&=& 2 B_0 \hat{m}+\frac{ 16 (2L_8-L_5) }{F_0^2} B_0^2 
(m_u^2+m_d^2)\nonumber\\
M_{88}^2&=&\frac{2}{3}B_0 (\hat{m}+2 m_s)+
\frac{16 (2L_8-L_5)  }{3 F_0^2} B_0^2 (m_u^2+m_d^2+4 m_s^2)\nonumber\\
M_{00}^2&=&M_0^2(1-\Lambda_1)+\frac{32(2 L_8-L_5) }{3F_0^2}
 B_0^2 (m_u^2+m_d^2+m_s^2)\nonumber\\
&+& \frac{2}{3}(1+\rho)B_0(m_u+m_d+m_s)\nonumber\\
M_{38}^2&=&\frac{1}{\sqrt{3}}B_0(m_u-m_d)+
\frac{16 (2 L_8-L_5)}{\sqrt{3}F_0^2} B_0^2 (m_u^2-m_d^2) \nonumber\\
M_{30}^2&=&-\sqrt{\frac{2}{3}}(1+\frac{\rho}{2}) B_0 (m_d-m_u)\nonumber\\
&+&\frac{16(2L_8-L_5) }{F_0^2}\sqrt{\frac{2}{3}} B_0^2 (m_u^2-m_d^2)\nonumber\\
M_{80}^2&=&\frac{\sqrt{2}}{3}(1+\frac{\rho}{2}) B_0(m_u+m_d-2 m_s)\nonumber\\
&+& \frac{16 (2L_8-L_5)}{F_0^2}\frac{\sqrt{2}}{3} B_0^2 (m_u^2+m_d^2-2 m_s^2),
\end{eqnarray}
where $\rho\equiv -\Lambda_1+2 \Lambda_2-8 L_5 \frac{M_0^2}{F_0^2}$.

The second class of NLO corrections can be grouped into a single term contained in the
${\cal{O}}(p^6)$ odd-intrinsic parity  WZ  Lagrangian\cite{Moussallam}---
\begin{eqnarray}
{\cal{L}}^{(6)\gamma\gamma}_{WZ}&=&-  i  \pi \alpha  t_1\langle\chi_{_{-}}
Q^2\rangle F\tilde{F}\nonumber\\
{\rm where }\quad\chi_{_{-}}&=&u^\dagger \chi u^\dagger-u \chi^\dagger u\quad{\rm with}\quad
u=\sqrt{U}
\end{eqnarray}
(There exists a second term \cite{Moussallam} that, upon the explicit inclusion of the
singlet pseudoscalar meson, becomes subleading in $1/N_c$ and is
therefore neglected.) The low energy constant $t_1$ has vanishing $\beta$-function  and its
value can be  estimated   by means of  a QCD sum rule for the 
general three-point function involving the
pseudoscalar density and two vector currents and  saturating the
spectral function in the hadronic sector with the states indicated in
Fig. (2), yielding\cite{Moussallam,Nyffeler}:
\begin{equation}
t_1=-\frac{1}{ m_V^4}(F_0^2+\frac{\tau}{M_{\pi'}^2}).
\end{equation}
Here the $F_0^2$ contribution is determined by the masses and decay constants of the
vector mesons (the vector meson mass is naturally taken to be
$m_V\simeq m_\rho$) and is represented by Fig. (2b),
while the contribution proportional to $\tau$ is determined by excited pseudoscalars,
 such as the $\pi'(1300)$, and is represented by Fig. (2a).  This latter 
contribution can be estimated within a model \cite{Fayya,Moussallam} and its
expected size is at most one third of the magnitude
of the vector meson contribution.  Since this is similar to the level
of uncertainty expected 
in the sum rule evaluation, the $\tau$-piece will be disregarded henceforth. 
As shown by the numerical analysis below, the effects on the $\pi^0$ width due to the 
   ${\cal{L}}^{(6)\gamma\gamma}_{WZ}$ with $t_1$ as estimated above are of similar 
magnitude to the rest of the NLO corrections and in the range of 0.5 \%.

\begin{figure*}
\includegraphics{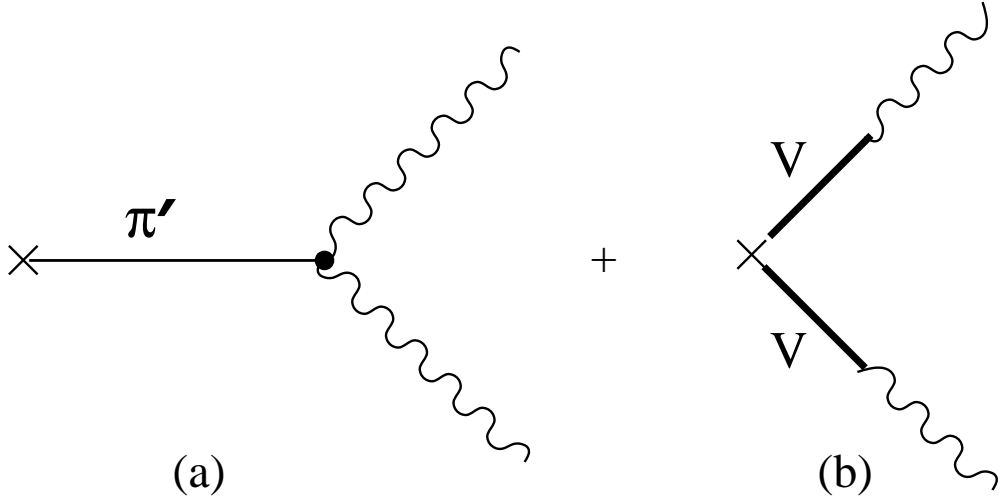}
\caption{\label{fig:widep} 
Hadronic contributions to $t_1$: $\pi'$  denotes excited pseudoscalar mesons and $V$ denotes vector mesons. }
\end{figure*}


At this point the  NLO two-photon amplitudes can  be explicitly given: 
\begin{equation}\label{NLOdecay}
\langle\gamma\gamma\mid \pi_{\bar{a}}\rangle=
-i\alpha\left(\frac{N_c}{12 \pi} C_a F^{-1}_{a\bar{a}} +
 \pi \frac{B_0}{F_0} t_1 \Theta_{\bar{a}a} \langle\{\lambda^a,{\cal{M}}_q\} Q^2\rangle\right)
\langle\gamma\gamma\mid F\tilde{F}\mid 0\rangle
\end{equation}
and the term proportional to $t_1$ can be obtained in two equivalent ways, either by
 determining the contribution from  ${\cal{L}}^{(6)}_{WZ}$
to the matrix elements  
$$\langle\gamma\gamma\mid \bar{q}\gamma_5\{\lambda^a,{\cal{M}}_
q\} q \mid 0\rangle$$
in Eq. \ref{div3}, or 
by directly calculating  the contribution to  $\langle\gamma\gamma\mid
 \pi_a\rangle$ due to that effective Lagrangian.  Note that the contribution from  ${\cal{L}}^{(6)}_{WZ}$ has the same scaling in $N_c$ as the leading one.  In Eq. \ref{NLOdecay} the factor 
$B_0 \langle\{\lambda^a,{\cal{M}}_q\} Q^2\rangle$ can be expressed using  the LO mass formulas, namely 
\begin{equation}
B_0 \langle\{\lambda^a,{\cal{M}}_q\} Q^2\rangle=\left\{
\begin{array}{c}
\frac{1}{9}(3 M_\pi^2+5(M_{K^+}^2-M_{K^0}^2)),~~~a=3 \\
\frac{1}{9\sqrt{3}}(7 M_\pi^2+M_{K^+}^2-5 M_{K^0}^2),~~~a=8\\
\frac{1}{9}\sqrt{\frac{8}{3}} (2 M_\pi^2+2M_{K^+}^2-M_{K^0}^2),~~~a=0
\end{array}
\right.
\end{equation}

At NLO the extraction of the ratio $R=m_s/(m_d-m_u)$ that characterizes the size of isospin breaking should be 
improved by including NLO corrections to Dashen's theorem. 
Over time several works have shown that the corrections are sizeable. 
From the early  works of Donoghue, Holstein and Wyler \cite{dhw} and 
Bijnens \cite{bij}, and more  recent works \cite{moreDashen}, it is  well
established that the mass difference $M_{K^0}-M_{K^+}$ left after subtracting the EM contribution 
is larger than the one predicted by Dashen's theorem. While Dashen's theorem predicts  $M_{K^0}-M_{K^+}=5.25~{\rm MeV}$, 
after the corrections have been implemented it is estimated that  $M_{K^0}-M_{K^+}=6.97 ~{\rm MeV}$. 
The mass difference is slightly smaller if the chiral logarithms are neglected following the approach of this work. 
In such a case    $M_{K^0}-M_{K^+}=6.47   ~{\rm MeV} $. These corrections to Dashen's theorem translate naturally into 
a smaller value for R, as shown in the analysis that follows. 
There is additional evidence that R is overestimated by applying Dashen's theorem, 
and that comes from the decay $\eta\to \pi^+ \pi^- \pi^0$. At lowest order the decay rate, which is proportional to 
$(m_d-m_u)^2$, is found to be a mere 66 eV \cite{sut}, which is almost a factor of four smaller than the experimental value of $281\pm 28$ eV. At NLO in the chiral expansion the rate is increased to $167\pm 50$ eV \cite{GandL1}, while dispersive analyses \cite{kam,lea} give $209\pm 56$ eV, which is still substantially below the experimental value. 
Clearly one way to make up for the difference is to increase 
$m_d-m_u$. Because of the large uncertainties in both the experimental and theoretical side  it is 
difficult to be precise, but it seems that the increment implied by the violations to Dashen's theorem  
mentioned above is in line with the enhancement required to explain
the observed  $\eta\to  \pi^+ \pi^- \pi^0$ width.  In principle a precise
measurement of the $\pi^0$ width could provide an independent
determination of $R$.  However, as shown by the results of the present
analysis, the $\pi^0$ width is affected by the corrections 
to Dashen's theorem only at the level of 0.5\%, 
which is unfortunately well below the experimental error of 1.4\%
 aimed at by PRIMEX and about the same as the 0.6\% uncertainty due to the 0.3\% error in the the experimental value of $F_{\pi^+}$ \cite{pdg}.


  There are nine low energy constants to  be  determined---$F_0$, $
B_0 m_i$ 
($i=u,d,s$), $M_0$, $L_5$, $L_8$, $\Lambda_1$, and $\Lambda_2$ and 
these can be found by solving for
the observables:
$F_{\pi^+}=92.42\pm 0.25~ {\rm MeV}$, $F_{K^+}=113.0\pm1.6~ {\rm MeV}$, $M_{\pi^0}=134.976~ {\rm MeV}$, $M_{\eta}=547.30 ~ {\rm MeV}$, $M_{\eta'}=957.78~ {\rm MeV}$,  $M_{K^0}=497.78~ {\rm MeV}$, $M_{K^0}-M_{K^+}$ (which as mentioned before, is 
5.25 MeV at LO, while  at NLO and disregarding (including) the chiral logarithms in the corrections 
to Dashen's theorem  is 6.47 MeV (6.97 MeV), $\Gamma_{\eta\to \gamma\gamma}=464\pm 45~{\rm eV}$, and  $\Gamma_{\eta'\to \gamma\gamma}=4.28\pm0.34~{\rm keV}$.   
Note that the tenth LEC $t_1$ cannot  at this stage  be extracted phenomenologically and thus  its value is taken  according to the estimate made above. 



In  Table I the second column displays the LO results, and the next three columns display  three different
 NLO fits, namely:
\begin{itemize} 
\item [i)] NLO\#1 includes  terms of ${\cal{O}}(p^6)$ and ${\cal{O}}(p^4/N_c)$
in the decay amplitude ---{\it i.e.} chiral logarithms are omitted.
\item [ii)] NLO\#2 includes
chiral logarithms, which are ${\cal{O}}(p^6/N_c)$, and the 
renormalization scale $\mu$ is set equal to $M_\eta$. 
\item [iii)] NLO\#3 is identical to NLO\#1 but sets
$t_1=0$ ---{\it i.e.} excludes the ${\cal{O}}(p^6)$ WZ contributions.
\end{itemize}

\begin{center}
\begin{table}
\caption{Results for the LECs for several different fits, the LO fit and three NLO fits as explained in the text. The LECs that depend of the QCD renormalization scale $\mu_{QCD}$ correspond to the limit
$\mu_{QCD}\to \infty$. Although all the LECs are the renormalized ones, only those in the NLO fit \#2 depend on the chiral renormalization $\mu$ that is  taken to be equal to $M_\eta$.}
\begin{tabular}{|c|c|c|c|c|}
\hline 
 LEC & LO Fit & NLO \#1 &   NLO \#2  & NLO \#3 \\
\hline
$F_0$ [MeV] & 92.5$\pm$0.6			& $90.73\pm 0.47$		& $84.54\pm 0.85$		&$90.69\pm 0.38$ 	\\
$M_0$ [MeV] & 848$\pm$40& $1047\pm 5$	& $1142\pm 33$	&$1044\pm 4$		\\
$2 B_0 \hat{m}$ [GeV$^2$] & $0.0366\pm 0.0001$	& $0.03656\pm 0.00001$	& $0.0362\pm 0.0002$	&$0.03648\pm 0.00001$\\
$B_0 (m_d-m_u)$ [GeV$^2$] 	& $ 0.0235\pm 0.0006$& $0.0237\pm 0.0002$& $0.0245\pm 0.0006$	&$0.0244\pm 0.0002$ \\
$B_0 m_s$ [GeV$^2$] 	& $0.236\pm 0.006)$& $0.2349\pm 0.0006$& $0.197\pm 0.002$ & $0.2291\pm 0.0005 $\\
$2 L_5+L_8$& 0& $(5.26\pm 0.01)\times 10^{-3}$	&$(6.3\pm 1.1)\times 10^{-3}$		&$(5.44\pm 0.07)\times 10^{-3}$\\
$2 L_8-L_5$     & 0	& $(0.8\pm 0.9)\times 10^{-5}$	&$(-0.53\pm 0.04)\times 10^{-3}$& $(0.11\pm 0.01)\times 10^{-3}$\\
$\Lambda_1$         	& 0	& $0.19\pm 0.01$		&$0.29\pm 0.04$				&$0.209\pm 0.006$\\
$\Lambda_2$         	& 0	& $0.74\pm 0.02$		& $1.4\pm 0.4$			& $0.81\pm 0.02$\\ \hline
\end{tabular}
\end{table}
\end{center}
\vspace{0.cm}
\begin{center}
\begin{table}
\caption{results implied by the different fits displayed in Table I. The $\star$ indicates that the quantities are inputs. }
\begin{tabular}{|c|c|c|c|c|}\cline{2-5}
 \multicolumn{1}{c|}{}
 & ~~~ LO Fit~~~ & ~~ NLO \#1 ~&
  ~ NLO \#2  ~ & ~ NLO \#3 ~ \\
\hline
$\Gamma_{\pi^0\to \gamma \gamma}$ [eV] & 8.08  & 8.10 &8.16 & 8.14\\
$\Gamma_{\eta\to \gamma\gamma}$ [eV]   &565 &464 $\star$ & 464 $\star$ & 464 $\star$ \\
$\Gamma_{\eta'\to\gamma\gamma}$ [keV] 	&5.1&4.28 $\star$ &4.28 $\star$ & 4.28 $\star$  \\
$M_{\pi^+}-M_{\pi^0}$ [MeV] 	&0.32 &0.24 & 0.16 $\star$ & 0.21\\
$m_s/\hat{m}$ 	& 25.9 & 25.7 & 21.7 &25.1 \\
$R=m_s/(m_d-m_u)$ & 45.3 & 36.6 & 30.9 & 37.5 \\
$\theta_3$ [$^o$]       &1.57 & 1.51 &1.88 &1.40\\
$\theta_8$  [$^o$]    	& -0.56& -0.68 & -0.94& -0.59\\
$\theta_0$  [$^o$]     	&-18.6 & -10.6 & -8.7 & -12.2\\ \hline
\end{tabular}
\end{table}
\end{center}

It should be noted that the mass of the $\eta'$ in the chiral limit
and at NLO in $1/N_c$ is given by $\sqrt{1-\Lambda_1} M_0 \sim 940$
MeV, which is slightly high leaving  not enough  room for the piece 
linear in the quark masses. This  linear contribution is suppressed by the
rather large value of $\Lambda_2$ which leads to a very small value of
$1+\rho$. This cancellation between the leading and subleading in $1/N_c$ contributions  seems to indicate some difficulty with the
$1/N_c$ expansion for the masses. Ther first manifestation of this problem is of course in the problem found with the $\eta$ and $\eta'$ masses at LO. Although this problem has a minor impact on $\pi^0$ width,  it certainly deserves further study. It
should be noted that the mass difference $M_{\pi^+}-M_{\pi^0}$ has
been given as input in fit \#2 since its value emerges as too large
if left unconstrained. The reason why it is too large can be traced 
back to the chiral logarithms generated by the $\eta'$. It seems,
therefore, that requiring the subleading renormalized LECs to vanish 
is not such a good approximation when such chiral logarithms are included.

Table II lists the associated predictions for various 
quantities of interest, in particular the $\pi^0$ width.
It is evident from the NLO fits that $\Gamma_{\pi^0\to \gamma\gamma}$ 
is rather  stable and  
always within $1\%$ of the leading order result. This is within the expected range
 of the NLO corrections. As shown by comparison of the first and third NLO fits,
the correction from the ${\cal{O}}(p^6)$ WZ Lagrangian 
${\cal{L}}^{(6)\gamma\gamma}_{WZ}$  reduces the  $\pi^0$ width by
$0.5\%$, 
a magnitude in line with the fact that such a correction  is 
controlled by the ratio $m_{u,d}/\Lambda_\chi$.  The chiral-logarithm 
contributions to the amplitudes as shown by fit \#2, provide an
increase of order $0.5\%$ to  the $\pi^0$ width. Since these are subleading 
corrections  of  ${\cal{O}}(p^6/N_c)$ to the decay amplitude, they  are somewhat larger  than the  $0.2-0.3\%$ expected from  the ratio 
$m_{u,d}/(N_c \Lambda_\chi)$ that determines them.  This problem is similar to the one  with the pion mass difference just mentioned; in this case the $\eta'$ loops affect the mixing angles producing a  larger than expected correction to the rate. Indeed,
 turning off the chiral logarithms generated by the $\eta'$ the $\pi^0$ width is essentially identical to  the result in fit \#1.


It is important to note that the mixing
angles are  substantially modified at NLO: in particular, the 
$\pi^0-\eta$ mixing angle is found to be $\epsilon\sim 
\theta_3+\theta_8=0.8^o-0.9^o$ in the
three fits, which is $\sim 10-20\%$ smaller than the  LO result of
$1^o$, but still larger than
that obtained at LO within $SU(3)$ and given in Eqn. \ref{SU3mixing}. 
The latter is chiefly  a consequence of the corrections to Dashen's 
theorem.  The $\pi^0-\eta'$ mixing $\tilde{\epsilon}$ goes from $0.5^o$ at LO to 
approximately $0.3^o$ at NLO;
finally, the $\eta-\eta'$ mixing angle is dramatically reduced 
to about $-10^o$ from its LO value of $-18.6^o$. In view of these substantial
corrections to the mixing angles, the stability of the $\pi^0$ 
width is nontrivial: besides the corrections due to the
${\cal{O}}(p^6)$ WZ Lagrangian, the decay amplitude is 
determined by the decay constant matrix $F_{\bar{a}a}$, which is 
affected by the mixing of states as well as by the NLO corrections 
contained in the decay constant matrix $F_{a b}$ given in
Eqn. \ref{NLOF}. It turns out that the entries in 
$F^{-1}_{a \bar{a}}$---namely $F^{-1}_{a \pi^0 }$---affecting 
the $\pi^0$ amplitude remain stable well within the natural size of the NLO corrections.

In order to asses  the theoretical uncertainty of  the analysis of the
$\pi^0$ width, an estimate of the magnitude of EM corrections should
also be given. Such corrections are of order $\alpha/2\pi$, which puts
them in the $0.2-0.3 \%$ range. Note also that the value of $F_{\pi}$ 
being used is that of $F_{\pi^+}$ which has an  EM correction.  
This correction has been studied \cite{Urech} and is given by  
$\delta_{\rm EM} F_\pi\sim 
\kappa 4 \pi \alpha F_0$ where the low energy constants that determine
the coefficient  $\kappa$ can be  estimated in a resonance saturation
model that gives $|\kappa| \sim 10^{-2}$, thus leading to the estimate
$|\delta_{\rm EM} F_\pi|\sim 0.1\; {\rm  MeV}$ which is well  within the experimental
uncertainty in the value of $F_{\pi^+}$, and implies a correction to the
$\pi^0$ width consistent with the range mentioned above.
 It seems, therefore, safe to neglect electromagnetic corrections at the level of precision  of this work. 

Although the $\eta-\eta'$ complex is not the primary focus of this work, the analysis carried out illuminates crucial aspects of this system. At LO the description is rather poor, in particular because the two-photon widths depart quite substantially from the experimental world averages. Indeed, $\Gamma^{LO}_{\eta\to \gamma\gamma}=613 \;{\rm eV}$ versus the world average
 experimental value of $464\pm 45\; {\rm eV}$, and
 $\Gamma^{LO}_{\eta'\to \gamma\gamma}=4.86\;{\rm keV}$ vs $4.28\pm
0.34\; {\rm keV}$. These latter disagreements are mostly due to the
large $\eta-\eta'$ mixing angle
that results  from the LO mass formulas, and the fact that at
LO all decay constants are set to be equal to $F_\pi$. At NLO the
scheme that emerges is the one already found in other works
 \cite{Kaiser1}, where the mixing angle of the pure $U(3)$
states is in the proximity of $-10^o$ rather than the $-20^o$ obtained
in LO, and where the decay constant matrix can be parametrized by means of two
angles and two decay constants \cite{Kroll,Kaiser1}.  
Following the conventions and notation in\cite{Kaiser1}, the present analysis 
gives (quantities are denoted in boldface not to be confused with
 quantities defined heretofore in the text):
${\bf F_0}\simeq 116 $ MeV, ${\bf F_8}\simeq 122$ MeV,  ${\bf \theta_8}\simeq -20^o$
${\bf \theta_0}=-2.5^o ~~{\rm to}~~0.5^o$,  and the angles  ${\bf \theta_0}$ and  ${\bf \theta_8}$ 
for the three NLO fits are respectively $(-0.9^o, -19.9^o)$, $(2.0^o, -19.0^o)$ and $(-2.5^o, -21.5^o)$ . The difference  ${\bf \theta_0}-{\bf \theta_8}$ turns out to be between $19^o$ and $21^o$, to be compared with the $19^o$ obtained in  \cite{Kaiser1}
(in a NLO estimate in that reference a value of $14^o$ is obtained,
which departs substantially from the one of the current
analysis). There 
exist numerous  studies of the $\eta-\eta'$ complex. It makes sense
only to compare results with those
using the two-angle scheme \cite{eta-etap}. Although some of the 
analyses in these references are purely phenomenological, these
results are in general in good  agreement with the results obtained in this  work.

  The quark mass  ratios obtained in the different fits deserve comment.
The ratio $m_s/\hat{m}$ is in good agreement with the standard value $24.3\pm 1.2$ obtained in SU(3) \cite{massratios}; in fit \#2 it  is a couple of standard deviations smaller, most likely because  the chiral logarithms included do not represent the full NNLO contributions. In all, this is not surprising as the assumption that the low energy constants that are subleading in $1/N_c$ can be disregarded is one of the important assumptions in the extraction of the standard ratios. However,
a comment is in order: it is observed that  using LO mass
formulas in the NLO results ---{\it i.e}., expressing
quark masses in terms of meson masses squared in the NLO terms --- leads
to a fit that is less stable and generates large corrections to the 
quark mass ratios. The ratio $R$ is smaller here than the standard value $42.3\pm4.5$, and this is simply because the corrections to Dashen's theorem have been included. The values for $R$ in fits \#1 and \#3 are about one standard deviation smaller than the standard value, while in \#2 the chiral logarithms involving the $\eta'$ loops give a substantial reduction (when these are turned off  $R$ is similar to the result in the other fits).
 An interesting observation is that setting $m_u=0$ leads to an inconsistent 
fit and a value for the $\pi^0$ width of 8.5 eV.

\section{Conclusions}

The decay rate for $\pi^0\rightarrow
\gamma\gamma$ has been calculated within a combined  chiral and $1/N_c$ expansion. 
At leading order in the expansion, the isospin-breaking induced mixing
of the pure $U(3)$ states increases the size of 
$\Gamma_{\pi^0\to \gamma \gamma}$ from the value $7.725$ eV predicted
by the lowest order chiral anomaly by more than 4.5\%. This effect is
largely due to the fact that the contributions from mixing with the
$\pi_8$ and $\pi_0$ add constructively, and are of similar 
magnitude. However, at LO the resulting $\eta,\eta'\rightarrow \gamma\gamma$
widths are found to be too large, and in general
the fit is quite poor.  There is a clear need then
for the NLO calculation, both in order to improve the results in the
$\eta-\eta'$ sector and to test the stability of the enhancement
of $\Gamma_{\pi^0\to \gamma \gamma}$ observed at LO.  The NLO calculation reveals
that the LO result for $\Gamma_{\pi^0\to \gamma \gamma}$ is quite robust, being
modified by the NLO corrections  by less than 1\%. This stability is, however, non-trivial.
Indeed, as already noted, at NLO the mixing angles are
substantially affected---the  mixing angles $\epsilon$ and $\tilde{\epsilon}$ 
are reduced by 10 to 30\% (a more dramatic reduction of roughly 50\%  results
for the  mixing angle
$\theta_0$). The $\pi^0$ width, however, is only slightly
affected because the effects that ultimately determine the corrections
to the amplitude are in the decay constant matrix
$F_{\bar{a}a}$ shown in Eq. 17.  This matrix is affected by the
mixing, 
and also receives NLO corrections that reside in $F_{ab}$,
and apparently the NLO modifications to the mixings are partially
compensated by the latter corrections in the case of the entries relevant to the $\pi^0$. 

The primary source of theoretical uncertainty 
in the present  calculation  of  $\Gamma_{\pi^0\to \gamma \gamma}$ resides in the value of $R$, which has an uncertainty of about 15\%.  Using the 
empirical formula resulting from the results in Table II, 
$\Gamma_{\pi^0\to \gamma \gamma}\sim(7.725+ 14.1/R)\;
{\rm eV}$, the uncertainty in $R$ translates into an error in the
$\pi^0$ width of $0.6 \%$.  Other sources of uncertainty are the NNLO corrections, of which the chiral-logs are an example, and which should be expected to
be in the range of $0.2-0.3\%$, and also EM corrections which
according to a straightforward estimate should be in a similar
range. All this together implies an uncertainty of less than  $1\%$ in the theoretical prediction of $\Gamma_{\pi^0\to \gamma \gamma}$. Note that 
there is an overall  uncertainty in $\Gamma_{\pi^0\to \gamma \gamma}$ due to the 0.3\% error in $F_{\pi^+}$. The fact that the theoretical prediction for  $\Gamma_{\pi^0\to \gamma \gamma}$ shows the 4.5\% enhancement that can be experimentally observed, and the fact that the experimental result with the smallest quoted error \cite{direct} lies more than three standard deviations below that prediction  lend great significance to the upcoming PRIMEX measurement.

It is evident from the above analysis that no predictions for the $\eta$ and
$\eta'$ two-photon widths can be made. Rather, these quantities are 
inputs, and their precise values do not affect in any dramatic way the
$\pi^0$ width. In a more complete study, wherein the analysis is
extended to additional processes such as $\eta\to \pi^+\pi^-\pi^0$, a more precise experimental knowledge of such widths would be necessary. A more extensive analysis would also illuminate the $1/N_c$ corrections encoded in the LECs $\Lambda_1$ and $\Lambda_2$ and help determine them more precisely. 

As this manuscript was being  completed an analysis by Ananthanarayan and Moussallam \cite{MoussallamNew} was posted where the electromagnetic corrections are studied in detail. Their result for the $\pi^0$ rate is  in agreement with the results in this work. Related  work is also being completed by Kaiser and Leutwyler \cite{KaiserNew}.

Acknowledgements:

We thank J.~F. Donoghue for his collaboration during the early stages of this work and for useful discussions.
One of us  (JLG)  would like to  especially  thank  H. Leutwyler for 
enlightening discussions, as well as  A. Gasparian,  J. Gasser, 
   P. Minkowski  and A. Nyffeler,  and
  the Institut f\"ur Theoretische Physik of the University of
Bern for the kind hospitality and support 
while part of this work was completed.
This work was partially supported by the National Science Foundation 
through grants \#~PHY-9733343 (JLG) and \#~PHY-9801875 (BRH),
by the Department of Energy through contracts  DE-FC02-94ER40818 (AMB) and  DE-AC05-84ER40150 (JLG),
and by  a  sabbatical leave support  from the
Southeastern Universities Research Association (JLG).


\end{document}